\documentclass{article}
\usepackage[T1]{fontenc}
\usepackage[utf8]{inputenc}
\usepackage{ismir} % Remove the "submission" option for camera-ready version
\usepackage{amsmath,cite}
\usepackage{amssymb}
\usepackage{graphicx}
\usepackage{color}
\usepackage[hyphens]{url}

\title{A Fourier Explanation of AI-music Artifacts} % analysis
\multauthor
  {Darius Afchar \hspace{1cm} Gabriel Meseguer-Brocal \hspace{1cm} Kamil Akesbi \hspace{1cm} Romain Hennequin}
  {
  Deezer Research, Paris, France\\
  {\tt research@deezer.com}
  }

\def\authorname{D. Afchar et al.} % , G. Meseguer-Brocal, K. Akesbi and R. Hennequin

\newcommand{\ie}{\textit{i.\,e.,} }
\newcommand{\eg}{\textit{e.\,g.,} }

\DeclareFontFamily{U}{wncy}{}
    \DeclareFontShape{U}{wncy}{m}{n}{<->wncyr10}{}
    \DeclareSymbolFont{mcy}{U}{wncy}{m}{n}
    \DeclareMathSymbol{\sha}{\mathord}{mcy}{"58} 

\begin{document}

\maketitle

\begin{abstract}
The rapid rise of generative AI has transformed music creation, with millions of users engaging in AI-generated music. Despite its popularity, concerns regarding copyright infringement, job displacement, and ethical implications have led to growing scrutiny and legal challenges. In parallel, AI-detection services have emerged, yet these systems remain largely opaque and privately controlled, mirroring the very issues they aim to address. This paper explores the fundamental properties of synthetic content and how it can be detected. Specifically, we analyze deconvolution modules commonly used in generative models and mathematically prove that their outputs exhibit systematic frequency artifacts -- manifesting as small yet distinctive spectral peaks. This phenomenon, related to the well-known checkerboard artifact, is shown to be inherent to a chosen model architecture rather than a consequence of training data or model weights. We validate our theoretical findings through extensive experiments on open-source models, as well as commercial AI-music generators such as Suno and Udio. We use these insights to propose a simple and interpretable detection criterion for AI-generated music. Despite its simplicity, our method achieves detection accuracy on par with deep learning-based approaches, surpassing 99\% accuracy on several scenarios.
\end{abstract}

\section{Introduction}
\label{sec:introduction}

\textit{``It's not really enjoyable to make music now [...] I think the majority of people don’t enjoy the majority of the time they spend making music.''} --- M. Shulman, CEO at Suno.

Meanwhile, millions of users seem to enjoy creating AI-generated music. As a result, it was recently reported that at least a fifth of music delivered to streaming platforms is now synthetic \cite{deezer_genAI}.
There is undoubtedly a hype around generative AI (\textit{GenAI}) as well as lots of investments made \cite{widder2024watching}.
But this view on musical creation and enthusiasm for GenAI is far from being a majority opinion \cite{sippy2024behind, suno_backlash}.
Lawsuits have been filed against several AI companies \cite{trial_suno_udio}.
Beyond the largely debated ethical implications and the many social risks
\cite{pelly2025mood, wei2024exploring, gebru2023artists, gautam2024melting, klincewicz2025slopaganda, klein2025provocations}, AI-music has specifically raised lots of question on copyright infringements \cite{goetze2024ai}.
It is estimated that 24\% of musicians' revenues are at risk in the next few years \cite{cisac_genAI}.

Most of these models are black box and owned by private companies. There is a lack of regulation to their development and use.
In parallel to this rise of GenAI, AI-detection services have started to emerge \cite{li2024audio}.
Nevertheless, these services are not faring better than GenAI in that they are also mostly black box and owned by private parties.
This begs the question: \textit{What is specific about GenAI? Why do AI-detectors work? What cues may they rely on?}

This paper may be seen as an interpretability study. We take a step back from the arms race between GenAI and AI-detectors and try to ask what is specific about synthetic content and how we might detect it. We analyze deconvolution modules that are common in generative models and mathematically prove that their output will produce small frequency artifacts -- \ie peaks (see Figure \ref{fig:encodec}).
This effect is related to the well-known \textit{checkerboard artifact} \cite{odena2016deconvolution}.
Unlike previous empirical observations, we provide an explanation for its origin using Fourier's framework (Section~\ref{sec:theory}).
Interestingly, we prove these artifacts depend not on the training data nor learned weights of generative models but on their chosen architecture.
In our experiments, we confirm the presence of this effect on multiple open-source models and that their structure aligns with what our theory predicts (Section~\ref{sec:exp}).
Further, we also show this effect happening with commercial AI-music generators (Suno and Udio).
In turn, we propose a simple and interpretable criterion to detect AI-music by exploiting this phenomenon.
Thanks to its underlying theory, its failure cases may be anticipated and explained.
Although providing a novel detector is not our main goal but a by-product of our analysis of synthetic artifacts, our detection scores are on par with deep learning-based solutions, with over 99\% accuracy.

\vspace{-0.8em}
\section{Related work}

\subsection{AI-music detection}

Mirroring the recent boom of commercial AI-music generation services (\eg Suno, Udio \cite{rahman2024sonics}), the task of AI-music detection is relatively novel. Only a few works have been published so far \cite{afchar2024detecting, li2024audio, rahman2024sonics}. These early works propose CNN-like models to learn to classify real and synthetic signals. They discuss several challenges, such as the difficulty of making these detectors robust to audio manipulations.
Nevertheless, a wealth of adjacent research topics may be related to this task.
Voice spoofing and synthetic singing voice detection models have been proposed \cite{wu2017asvspoof, almutairi2022review, sun2023ai, zang2024singfake}.
The literature on the detection of other modalities of synthetic media is also vast: \eg regarding video deepfakes \cite{mirsky2021creation}, or generated text \cite{lin2024detecting}.
In all these works, the dominating approach is to learn to classify synthetic and real signals with black box models and then focus on their robustness. Instead, our work rather tries to explain why synthetic samples may be detected and what cues models may rely on.

\vspace{-0.4em}
\subsection{Checkerboard artifact}

This phenomenon was first analyzed in \cite{odena2016deconvolution}. It was argued to be linked to overlaps of deconvolution kernels. It was also largely analyzed in computer vision for image synthesis and its detection (\eg \cite{zhang2019detecting, osakabe2021cyclegan, liu2022detecting, corvi2023intriguing}). In audio, this effect was first noticed by \cite{donahue2018adversarial}, which noted that it led to audible "pitched noise". In \textit{MelGAN} \cite{kumar2019melgan}, convolution hyperparameters were argued to be chosen properly to avoid this overlap and artifact. \cite{pons2021upsampling} further recommends using interpolating-upsampler to reduce the effect. 
Our main analysis is similar to that found in these latter works.
However, while most existing work regards videos and speech, our focus on AI-music is novel and timely (\eg Suno and Udio).
Then, our analysis and modeling of deconvolution artifacts in the frequency domain, as well as their usage for detection with a conveniently defined, simple linear model instead of a black box model, are not frequently found in the literature.
Finally, to our knowledge, our analysis of the \textit{architecture-dependence} and \textit{training-} and \textit{data-independence} had not been discussed before.

\vspace{-0.2em}
\section{Fourier analysis of artifacts}
\label{sec:theory}

In this work, we propose to reinterpret Convolutional Neural Networks (\textit{CNN}) under the lenses of Fourier transforms. Instead of viewing inputs and outputs and hidden layers as time-based signals, we analyze their frequency-based decompositions (\ie spectra).

In the field of MIR, it is frequent to use spectrograms to process audio signals, as they better align with the human ear's perception. It is less common to do the same operations for everything happening within generation models.
Nevertheless, this framing has many natural advantages to interpreting CNNs. For instance, a convolution in the time domain is dual to a multiplication in the frequency domain.
This means a CNN may be reinterpreted as performing a series of multiplications.
This property, as well as others, will be informative to better explain the emergence of generation artifacts.

Our overall proof sketch is to show that the deconvolution operation periodizes the spectra of hidden layers, hence creating peaks by tiling the constant component of the signal. Then, we explain that this property persists through the following model's operations and layers.
Interestingly, our theory suggests that this phenomenon should only depend on a chosen model architecture but not on the training weight or data. We confirm this property in our next experiment section.

\begin{figure*}[t!]
  \centering
  \includegraphics[width=0.9\linewidth]{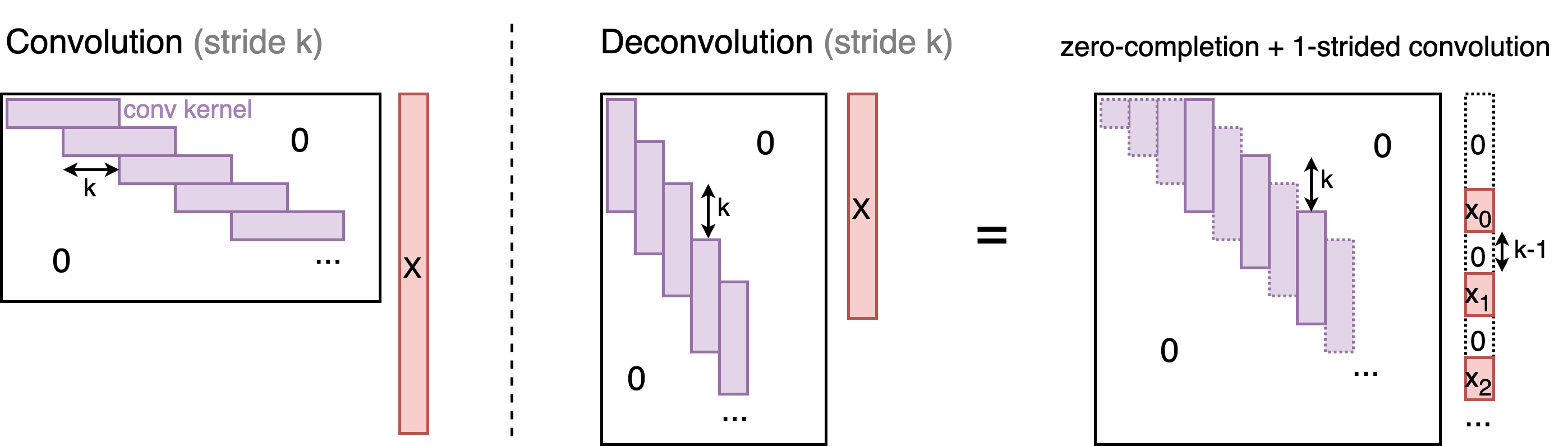}
  \caption{A CNN's convolution and deconvolution of an input $X$ may be expressed as matrix multiplications. To build these matrices, the kernel is tiled along the diagonal with steps $k$ (the stride). Then, it may be verified that a deconvolution is equivalent to a 1-strided convolution of the input with corresponding $k-$1 zeros inserted between each value.}
  \label{fig:deconvolution}
  \vspace{-0.5em}
\end{figure*}

\subsection{Background}

We first report some essential concepts and properties to better understand our following arguments and proof.

\vspace{1em}
\textbf{Fourier transform.}
We denote this transform $\mathcal{F}$. For a given integrable signal $s(t)$,
% supprimable
$ \mathcal{F}[s](\xi) = \int _{-\infty }^{\infty }s(t)\ e^{-i2\pi \xi t}\,dt $.
The transform is linear in $s$. 
Any multiplication (denoted~$\cdot$) is transformed into a convolution (denoted~$*$), and vice-versa: $\mathcal{F}[s * r] = \mathcal{F}[s] \cdot \mathcal{F}[r]$.
As a reminder, a convolution is computed as $(s * r)(t) = \int_{-\infty }^{\infty } s(\tau)r(t-\tau)d\tau$.
We recommend \cite{fourier_mallat} for an extensive presentation of the Fourier transform.

\vspace{1em}
\textbf{Dirac.}
A \textit{Dirac} distribution is an impulse function denoted as $\delta_x$, with a measure $\| \delta_x \|_1 = 1$, and equal to 0 everywhere except in its parameter $x$. A property we will use is that a convolution with a Dirac is equivalent to applying an offset to a function: $(s * \delta_x)(t) = s(t-x)$. 

\vspace{1em}
\textbf{Dirac comb.}
A \textit{Dirac comb}, denoted as $\sha_T$, is defined as a sum of Dirac impulses evenly spaced with a period $T$:
$\sha_T = \sum_{n = -\infty}^{\infty} \delta_{nT}$.
The Fourier transform of a Dirac comb is also a Dirac comb, but with spacing $1/T$: $\mathcal{F}\left[ \sha_T \right] = 1/T \, \sha_{1/T}$.

Dirac combs are useful to formalize the concept of \textit{sampling}. Multiplying a continuous signal with a Dirac comb leads to a discretized version of that signal.
Interpreted in the frequency domain, the Fourier transform of that discretized signal is equal to the convolution of the continuous spectrum with a Dirac comb of period $1/T$.
By distributing the sum of the comb, this equals a sum of Dirac convolutions, which as we have just mentioned, results in a sum of several offsetted versions of the spectrum:
$$\mathcal{F}\left[ s \cdot \sha_T \right](\xi) = \frac 1 T \sum\limits_{n = -\infty}^{\infty} \mathcal{F}\left[ s \right](\xi - n/T)$$
This phenomenon is called a \textit{periodic summation}.

In Nyquist-Shanon sampling theory, this concept is used to explain the phenomenon of \textit{aliasing}: \ie when copies of the spectrum will overlap, and why to chose the sampling frequency $1/T$ higher than twice the spectral bandwidth of a signal $s$ to enable a perfect reconstruction of the sampled signal.
In the next section, we use this phenomenon slightly differently to explain that this periodization is happening within deconvolution layers.

\vspace{1em}
\textbf{Deconvolution.}
Many generation models rely on ``deconvolution'' modules in their architectures. % \footnote{This concept is distinct from the deconvolution in signal processing.}
A deconvolution mirrors the way a strided convolution layer operates.
Instead of shrinking an input into a latent representation, it expands it.
It tiles a parametric vector (the \textit{kernel}), which is multiplied by each coordinate of the input, this with a striding factor \cite{dumoulin2016guide}.
Deconvolutions are also sometimes called ``transposed convolutions''. Indeed, a strided convolution can be rewritten as a multiplication with a big matrix where the learned kernel is tiled horizontally along a diagonal with a displacement equal to the stride (Fig. \ref{fig:deconvolution}). With this view in mind, it may be verified that the deconvolution -- as defined in CNNs -- corresponds to multiplying an input with a matrix with a similar but transposed structure: \ie the kernel tiled vertically with a displacement equal to the sought upsampling factor (Fig. \ref{fig:deconvolution}).

Then, an important property is to realize that a strided deconvolution can be rewritten as the composition of an upsampling operation with zeros filling in-between values, followed by a 1-strided convolution. An illustration of the proof of this property is provided in Figure \ref{fig:deconvolution}. For the complete formal proof, we refer interested readers to \cite{dumoulin2016guide}. % [Sec. 4.5]

As reported in \cite{odena2016deconvolution}, there exist alternative definitions of deconvolution layers: an interpolated upsampling operation (\eg linear, bicubic) followed by a convolution. We will discuss this formulation later and keep the first definition of deconvolution modules for now.

\subsection{Deconvolutions induce periodization}

We are now ready to formulate our main analysis.
As we have reported, deconvolution can be seen as performing an upsampling operation with zero-insertions between values, followed by a convolution.
Let us show that this zero-insertion leads to a periodization of the signal.

\begin{figure}[h!]
  \centering
  \includegraphics[width=\linewidth]{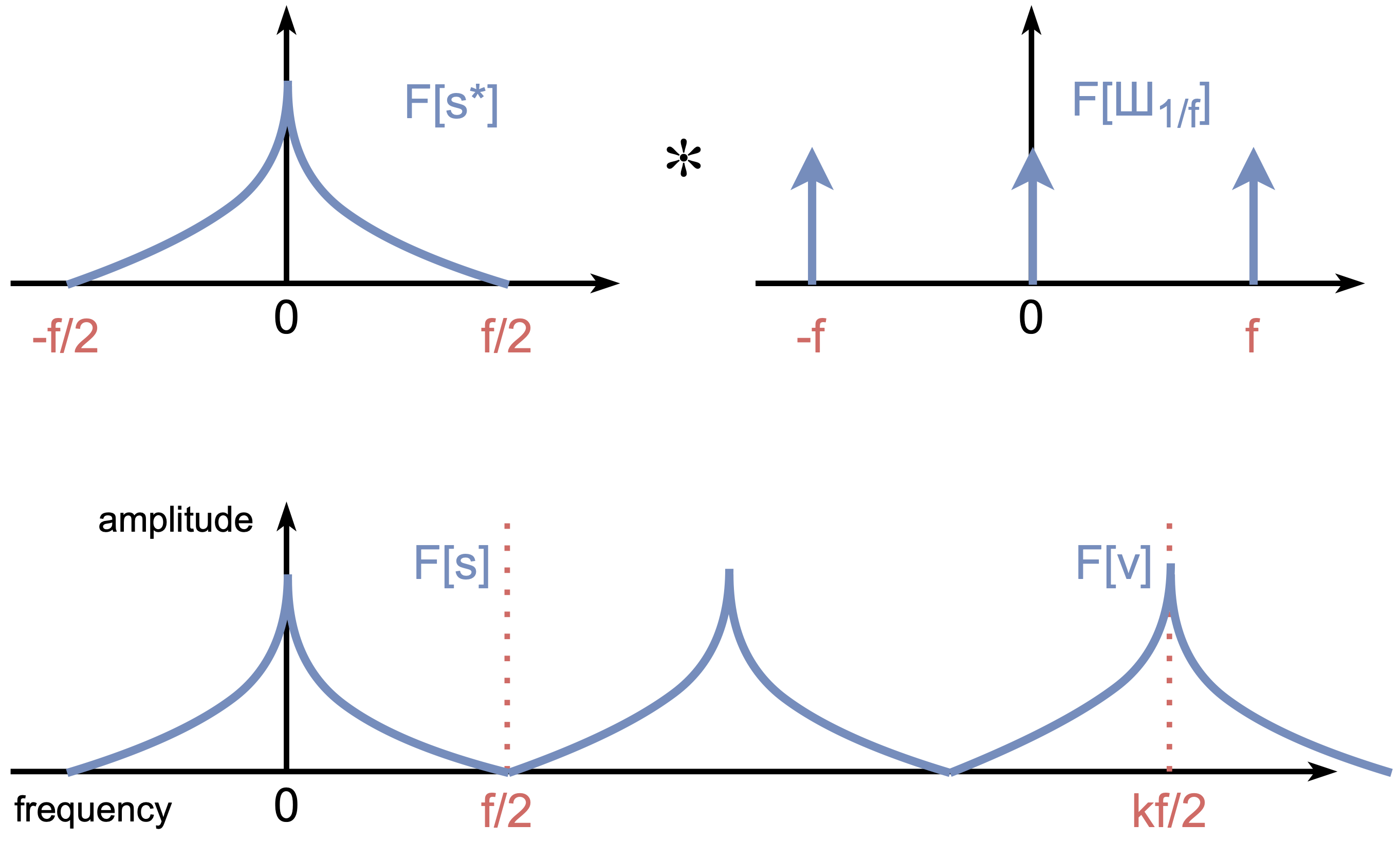}
  \caption{Periodization due to the zero-upsampling. Dotted red lines indicate the discrete signals' sampling cutoffs.}
  \label{fig:periodization}
  \vspace{-1em}
\end{figure}

Zero-upsampling is equivalent to over-sampling a discretized signal with a multiple of its sampling frequency.
Let us consider a signal $s$, discretized with a sampling frequency $f_s$, from a continuous signal $s^*$: % with a bandwidth $f/2$
$$s = s^* \cdot \sha_{1/f_s}$$
Considering a deconvolution with stride $k$, 
the zero-upsampled version $v$ of $s$ may be interpreted as over-sampling $s$ with a multiple $k$ of the frequency of $f_s$:
$$v = s \cdot \sha_{1/kf_s} = s^* \cdot \sha_{1/f_s} \cdot \sha_{1/kf_s} = s^* \cdot \sha_{1/f_s}$$
As consequence, from a continuous transform perspective, $s$ and $v$ have exactly the sample spectrum: $\mathcal{F}[s] = \mathcal{F}[v]$. However, moving to the discrete transform, while $\mathcal{F}[s]$ is read up to its proper sampling frequency $f_s$, $\,\mathcal{F}[v]$ is read up to a frequency $kf_s$, as shown in Figure~\ref{fig:periodization}.
Thus, the zero-inserted signal $v$ contains multiple clones of the spectrum of $s$ because of the periodization effect shown above.

This is interesting since, empirically, common spectra of music signals often have most of their energy around the 0 frequency (\ie their mean value) and an exponential decrease of energy in higher frequencies. Said otherwise, music spectra look like skewed triangles\footnote{Note: for real signals, due to the conjugate symmetry of the Fourier transform, the magnitude spectrum is mirrored for negative values of $\xi$.}.
Furthermore, in hidden layers, it is common to add biases in the output of learnable layers, or use ReLU activations. These operations further create a bias in the time domain that leads to a "peak" at $\xi = 0\textrm{Hz}$ in the frequency domain.
Therefore, when a signal (or hidden layer vector) is periodized, this triangle pattern and its peak in 0 are cloned throughout the spectrum.
We claim that this periodization of the high-energy bias is what may lead to checkerboard artifacts \cite{odena2016deconvolution}. In music, this effect translates as a hissing noise \cite{donahue2018adversarial, kumar2019melgan}.

To formalize the replication, peaks in the spectrum may be found at each $f_s$ period up to the sampling frequency $kf_s/2$ of $v$. Thus, let $\,p_{\max} = \lfloor k/2 \rfloor$, we have peaks for all integer $n \in [ 0 \, .. \, p_{max}]$ at the frequencies $nf_s$. For instance, a deconvolution with stride 8 leads to 5 peaks.

\vspace{1em}
We have seen that the zero-upsampling of a deconvolution changes the expected distribution of the audio spectrum, from a general triangle shape, to a concatenation of triangles and peaks. How is this shape impacted by the following convolution of the layer?

As reminded, a convolution in the time domain is equivalent to a multiplication in the frequency domain. 
Coming back to our peaked triangles and reading this from an amplitude spectrum in log-scale: a convolution is equivalent to a vertical offset in the log-spectrum.
Importantly, there is a big difference between the typical size of commonly processed music signals (\eg 1s of audio may contain 48000 samples), and the size of CNNs' kernels (\eg 3, 5 or 7).
In the frequency domain, the space of possible spectra the kernel's transform lives in is parametrized by only a few parameters\footnote{which is the size of the kernel: \eg the discrete transform of a kernel $K$ of size three is $\mathcal{F}[K](\xi) = k_0 + k_1 e^{-2i\pi \xi/N} + k_2 e^{-4i\pi \xi/N}, \; \forall \xi \in [0, N-1] $}.
To multiply these two spectra, the kernel is zero-padded, which is equivalent to a \textit{spectral interpolation} \cite{fourier_mallat}.
Said otherwise, the kernel spectrum is typically a slowly evolving function -- relative to the variations of the spectrum to be multiplied with.
While some very particular choices of kernels could mitigate the peaks (\eg a constant kernel of size the stride would cancel all peaks but the one at 0Hz), the limited expressivity due to the limited amount of parameters of the kernel is unlikely to remove all those artifacts entirely. As we will see in the next sections, this hypothesis is confirmed in practice: the output of a deconvolution actually result in creating peaks that are preserved after the convolution layer. Examples of such outputs may be found in Figure \ref{fig:encodec}.

% can be deleted if I lack some space
\vspace{1em}
Finally, we have alluded to the fact that some variants of deconvolution may include interpolations instead of the zero-upsampling we discussed.
The extension of our result to the general case of \textit{any} possible interpolation scheme is not straightforward. Still, there is one case that can be easily extended: linear interpolation.
For this, we remark that a linear interpolation may be computed as a convolution of the zero-upsampled input by a triangular filter $\Lambda(t)$ of width $k$, the stride of the deconvolution\footnote{With the linear interpolation, $\mathcal{F}[\Lambda](\xi) =k \, \textrm{sinc}^2(k \xi)$. The interpolation acts as a low-pass filter. This is why this variant is less frequently used as it may lead to worse reconstructions of high frequencies.}.
This convolution may be absorbed into the following convolution of the layer, and our results remain unchanged.

\subsection{Effect of multiple layers}

We have seen that deconvolutions, by nature of the zero-upsampling they induce, create artifacts due to the replication of the peak created by the input bias. Is this property preserved through multiple layers?
It is hard to give a general answer due to the variety of architectures that exist. We discuss the case of simple, sequential, CNNs.

\begin{figure}[h!]
  \centering
  \includegraphics[width=0.91\linewidth]{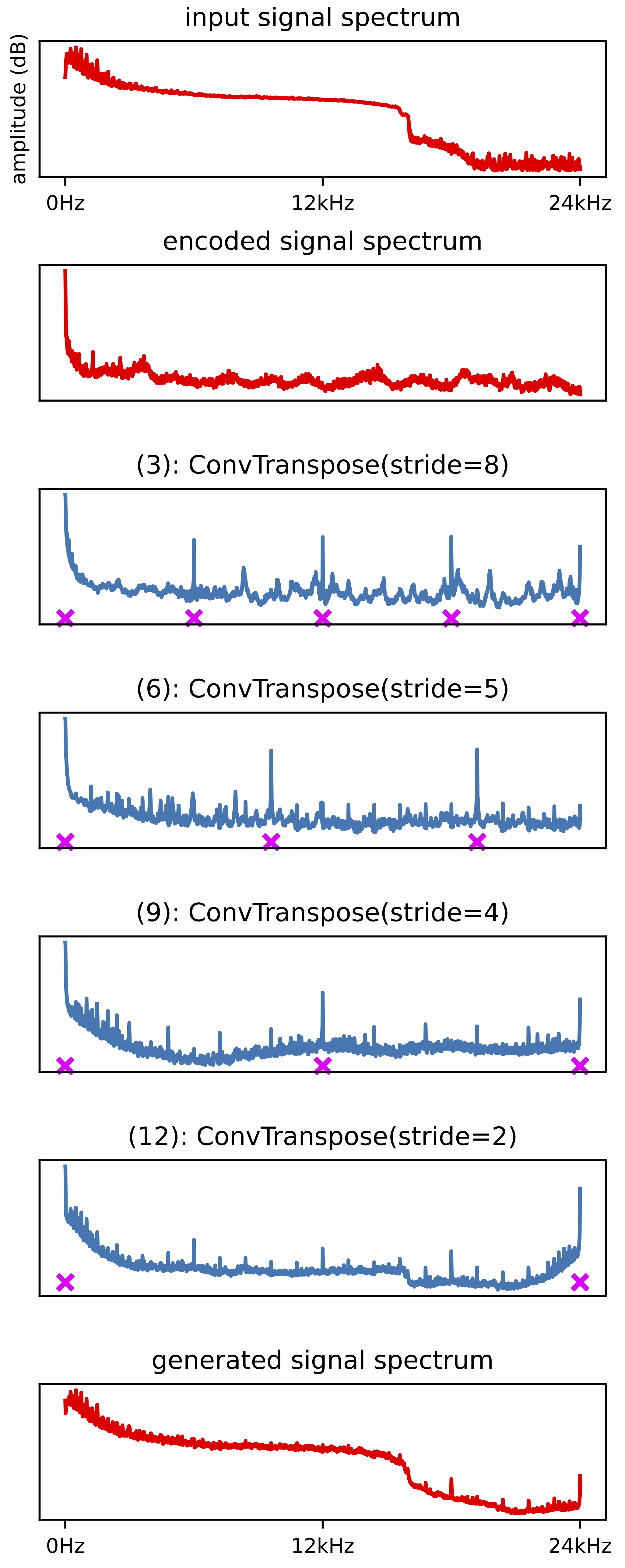}
  \caption{\textbf{Inside Encodec's decoder.} We plot the average spectra of a real music input, its latent representation (\ie after the dequantization of the audio tokens), as well as all its successive hidden outputs of all the deconvolution layers of Encodec (48kHz version). We indicate the layer number and chosen stride hyper-parameter of each deconvolution. We plot crosses below the signal where the periodized continuous component might create an artifact due to the periodization of each above spectrum (as deduced from the stride). The spectra are averages from the many channels of each hidden layer, as well as temporally averaged from successive $2^{13}$ samples.}
  \label{fig:encodec}
\end{figure}

In this case, by induction, the periodization induced by deconvolution may lead to replicating not only the continuous component peak but also all previous peaks. This means that artifacts will be cloned in a fractal-like manner.
To illustrate this property, we have computed the average spectrum found at several stages of the generator part of the Encodec model \cite{Defossez2022HighFNEncodec} in Figure \ref{fig:encodec}.
We can see peaks appearing at the predicted places (see the formula from the previous section), as well as preservation of previous layers' peaks in a fractal-like manner.

We formalize this recursive pattern. 
Instead of counting replications of the spectrum, it is convenient to count clones of the half-spectrum above 0 (the negative spectrum being symmetric for real signals).
Indeed, we have seen that a spectrum from a layer $i$ will be cloned $k^{(i+1)}/2$ times by a deconvolution $i+1$, thus creating $k^{(i+1)}$ half-spectra. By induction, a single peak at $0$Hz in the input will thus lead to $P_{\max} = \prod_{i=0}^L k^{(i)}$ cloned half-spectra after $L$ deconvolutions. In total, this amounts to $\lfloor P_{\max}/2 \rfloor + 1$ peaks when assembling each half-spectrum. For instance, Encodec has strides $\{ 8, 5, 4, 2 \}$ and thus creates 161 peaks.

% We formalize this recursive pattern. By induction, if there exists a set of peaks $\{ nf_s \mid n \in [0 \, .. \, p_{\max}^{i}] \}$, for a layer $i$, then a deconvolution $i+1$ with a stride $k^{i+1}$ will lead to a set of peaks $\{ nf_s \mid n \in [0 \, .. \, p_{\max}^{i+1}] \}$, defined with $p_{\max}^{i+1} = (\lfloor k^{i+1}/2 \rfloor + 1) p_{\max}^{i}$. For instance, from a single peak at $0$Hz in the input, $L$ deconvolution layers with a stride 2 may result in $2^L$ evenly spaced peaks in the output.

% supprimable si manque de place
% We remark that the periodization also clones negative values, which is why a $\textrm{stride}=8$ deconvolution leads to 5 peaks: what must be counted as 8 clones is each spectrum replicas plus each negative value conjugate. We report to Figure \ref{fig:periodization} for an example to reason with.

Interestingly, this recursive property suggests that if artifacts from previous layers are well preserved, then they may be used as a fingerprint of the whole architecture used by a generation model (specifically, the stride hyperparameter of each deconvolution).

\subsection{Discussion}

By combining several results from signal processing and properties of deconvolution, we have shown that generative CNNs may be subject to the generation of peaking artifacts in the spectra of their outputs. 
Our discussion has only regarded simple cases and does not cover all possible model architectures.
For instance, we have not discussed whether non-linear activations allow these artifacts to persist, nor whether skip connections or batch normalization may help reduce the bias peak and, hence, their replication in the spectrum.
In the next section, we will address these further questions empirically by analyzing the artifacts of several music generation models.

% Another aspect to discuss is that deconvolution layers do not only have one but several channels. Does the combination of several channels -- \eg by a following convolution -- impacts the presence of artifacts? Mostly likely not, because the deconvolution is the same for all channels, the periodization and artifacts placement is the same in all channels. Let $s$ be a signal that contains periodization artifacts, and $[K_1, ... K_N]$ a collection a convolution kernels. Any linear combination $(a_1, ... a_N)$ of the channels results in a spectrum $$\mathcal{F}\left[\sum\limits_{n=1}^N a_n(s * K_n)\right] = \mathcal{F}[s] \cdot \left( \sum\limits_{n=1}^N a_n\mathcal{F}[K_n] \right)$$
% by linearity of the transform.

Interestingly, we did not need to consider training data or model weights to show the emergence of artifacts. This suggests that this issue seems not related to nor solvable with better training and data, but is inherent to the use of deconvolution layers in a model.

\section{Experiment}
\label{sec:exp}

In this section, we validate our theoretical findings and close the gaps left out by theory with empirical results. 
Our research questions are the following:
\begin{itemize}%[leftmargin=0mm]
    \item[\textit{RQ1}] Are the artifacts solely architecture dependent?
    \item[\textit{RQ2}] Can they be used to detect synthetic music?
\end{itemize}
We show that the models we study (both open-source and closed-source) exhibit peaks in their generated outputs.
In turn, as an application example, this can be used to build a simple but efficient synthetic music detector that is both fast and interpretable.

\subsection{Setup}

We briefly detail the models and datasets we leverage in our experiments. More implementation details may be found on our code repository\footnote{\url{github.com/deezer/ismir25-ai-music-detector}}.

\vspace{0.5em}
\textbf{Open-source models.}
We consider several music generation models.
There are not many studies on AI-music detection.
To compare our results to \cite{afchar2024detecting}, we consider the same models: \textit{DAC}, \textit{Encodec} and \textit{Musika!}, as well as the medium split of the \textit{FMA} dataset \cite{Defossez2022HighFNEncodec, kumar2024high, pasini2022musika, fma_dataset}. The open-source nature of these models allows inspection.
As argued in \cite{afchar2024detecting}, many AI-music models rely on neural codecs (\eg \cite{garcia2023vampnet, copet2024simple}). This means that instead of generating a song end-to-end, it is more efficient to learn to generate a sequence of audio tokens, that are then dequantized and converted to 48kHz audio through the neural codec decoder. It was shown that learning to detect neural codec then transfers to the detection of music generators using that codec \cite{afchar2024detecting}.
We will compare our results to the black box detection model proposed in that latter study.
To address \textit{RQ1}, we also consider the \textit{MTAT} and \textit{MTG-Jamendo} datasets \cite{law2009evaluation, bogdanov2019mtg}.

\vspace{0.5em}
\textbf{Closed-source models.}
We also conduct experiments on \textit{Suno} and \textit{Udio}. We leverage the recent \textit{SONICS} dataset \cite{rahman2024sonics}, containing 50k of such synthetic music tracks.
With these models, we cannot verify whether the presence of peaks corresponds to the underlying architecture. We only showcase that these models exhibit the same issues.

\vspace{0.5em}
\textbf{Artifact fingerprint.}
To study the presence of artifacts, we need to extract the potential replicated peaks in generated spectrograms.
With the analysis provided in Section \ref{sec:theory}, it seems relevant to start from average spectra, as proposed in Figure \ref{fig:encodec}, and related to what was proposed in computer vision in \cite{corvi2023intriguing}.
Indeed, while a local music patch will exhibit many frequencies due to its melodic content, it is reasonable to think that when averaging many patches' spectrum over several minutes of audio, the melodic content will be smoothed and result in the triangle-like shape we were discussing. From there, we simply propose to subtract the local minima of the spectrum over sliding windows to highlight the local variations we try to detect (\ie peaks).
Finally, we consider a reduced bandwidth (\eg [5kHz, 16kHz]) to further discard melodic information and uninformative noise above the cutoff of the mp3 codec.
We refer to this small processing of the spectrum as computing an \textit{artifact fingerprint}. It represents the local variations in the amplitude of the average spectrum (\eg see Figure \ref{fig:dac}).

\subsection{Architecture dependence}

\begin{figure}[h]
  \centering
  \includegraphics[width=0.9\linewidth]{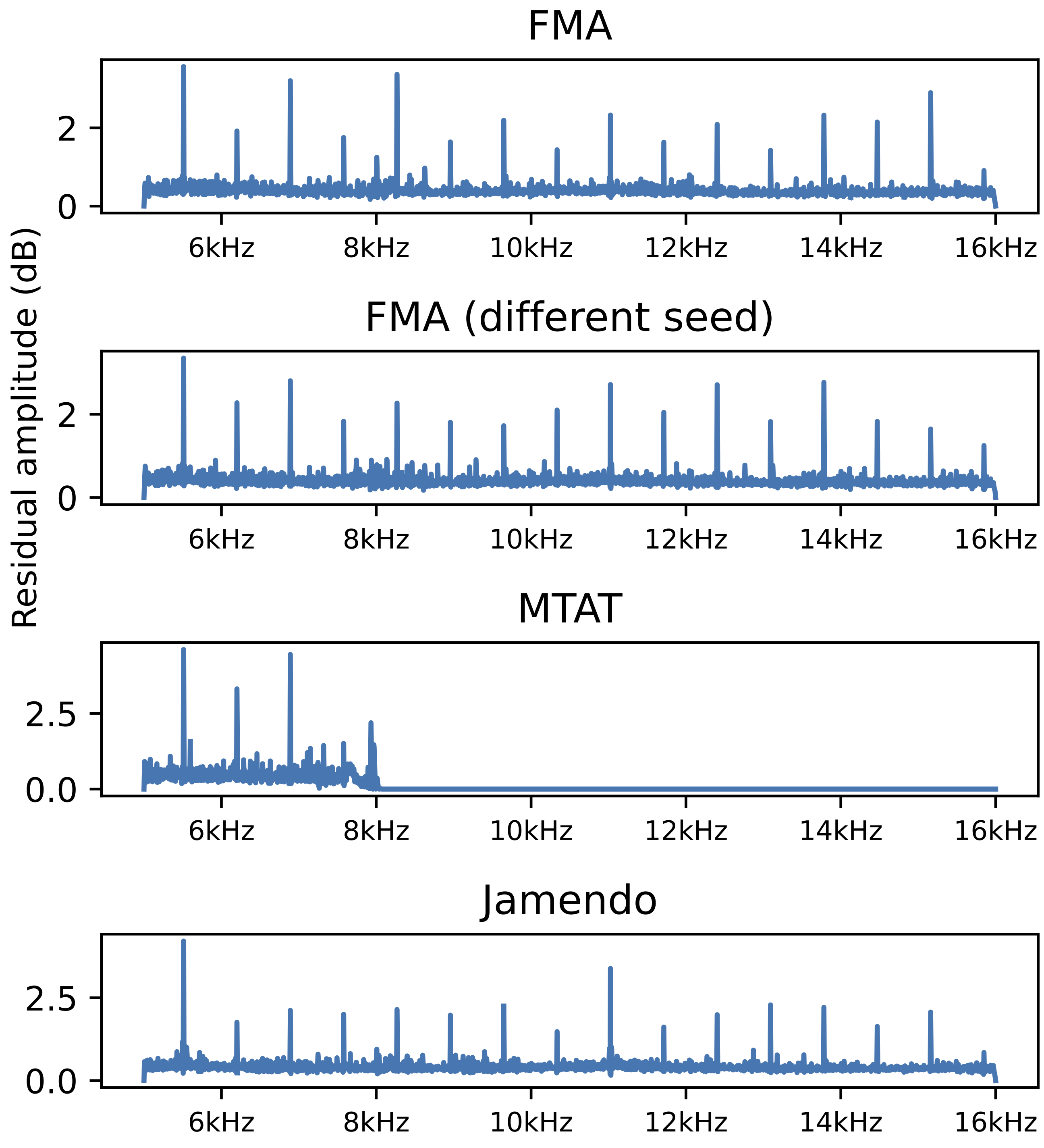}
  \caption{Artifacts of various trained versions of DAC.}
  \label{fig:dac}
  \vspace{-1em}
\end{figure}

To address \textit{RQ1} that the studied artifacts are independent of training data and learned weights, we train several models of \textit{DAC} \cite{kumar2024high}, using the same model configuration as \textit{VampNet} \cite{garcia2023vampnet}.
We train DAC on the FMA twice, with a different random seed (impacting the weight initialization and optimization), on MTAT and MTG-Jamendo. We compute artifact fingerprints of the auto-encoded tracks (following the same methodology as \cite{afchar2024detecting}). The averages of the found artifact fingerprints (on the test set) are displayed in Figure \ref{fig:dac}.
Strikingly, we find that the placement of peaks is the same for all four models.
Their amplitudes are slightly different between the four versions but still clearly exhibit the same overall pattern.
This means that artifacts are indeed: 1) \textit{Weights-independent}, since different training seeds do not change the peaks on FMA; 2) \textit{Data-independent}, since we find the same peaks for models trained on MTAT and Jamendo. Note that tracks from MTAT have a sampling rate of 16kHz, hence the cutoff that we observe at 8kHz.
Overall, this suggests that AI-music artifacts are indeed solely \textit{architecture dependent}.

This also suggests that a detector trained on one given model and relying on these cues may transfer its performance to that model trained on a different dataset, or even other models that share a similar parametrization of their deconvolution layers. This may explain the intra-family generalization property found in \cite{afchar2024detecting}.

\subsection{AI-music detection}

To address \textit{RQ2}, we build a simple detector.
From our previous section, it seems that synthetic content may be simply detected by checking if the artifact fingerprint contains peaks or not.
By the look of what was found in Figure \ref{fig:dac}, it seems that a straightforward linear regressor could be enough for this task: \eg learning positive coefficients for peaking frequencies and negative ones for baseline values.

We first follow the setup of \cite{afchar2024detecting}. We train to detect each real music track in FMA against its synthetic reconstructions. The results are provided in Table~\ref{tab:opensource}. We use the neural codecs at their maximum bandwidth, which is the hardest to detect.
Overall, the performances are comparable between our linear model and the reported black box CNN model, and the classification is almost perfect.

\begin{table}[h]
    \small
    \centering
    \begin{tabular}{|l|c|c|}
        \hline
        Class & \textbf{Our} & \textit{Reported from \cite{afchar2024detecting}} \\
        \hline
        \textbf{Real} & 99.87 & 99.7 \\
        %Synthetic (all) & 99.82 & 99.9 \\
        % $\hookrightarrow$
        %\hline
        \textbf{Synthetic} &  & \\
        $\hookrightarrow$ DAC (14kbps) & 99.68 & 99.3 \\
        $\hookrightarrow$ Encodec (24kbps) & 99.81 & 99.7 \\
        $\hookrightarrow$ Musika! & 99.97 & 100.0 \\
        \hline
    \end{tabular}
    \caption{Test detection scores (\%) on open-source models. We report the synthetic class breakdown.}
    \label{tab:opensource}
    \vspace{-0.5em}
\end{table}

Next, we do the same experiment with closed-source generators. We follow the setup of SONICS \cite{rahman2024sonics}. However, the real audio tracks were not provided due to copyrights. Therefore, we fall back to using tracks from FMA instead. We resampled them to 16kHz as \cite{rahman2024sonics}. This means our transforms have a cutoff at 8kHz. We accordingly adjust the bandwidth of our fingerprints to [1kHz, 8kHz].
This change, unfortunately, means that we cannot compare our results in a fair manner.
We nevertheless report SONICS' best model (\textit{SpecTTTra-$\alpha$}) in Table \ref{tab:closedsource} for information.
As it may be seen, the scores of our 10K-parameter logistic regression are comparable to the ones of a 20M-parameter transformer model.
On the synthetic version seen during training (\textit{Suno v3.5} and \textit{Udio 130}), the classification is perfect.
The performance drops on \textit{Udio 32}, which was unseen during training (following the splits from \cite{rahman2024sonics}). We suspect that Udio might have changed their model architecture between the 32 and 130 versions, which would explain the failure at the zero-shot performance transfer.

\begin{table}[h]
    \small
    \centering
    \begin{tabular}{|l|c|c|}
        \hline
         Class & Our & \textit{Reported from \cite{rahman2024sonics}} \\
         \hline
         \textbf{Real} & 99.97 & 99 \\
         %Synthetic (all) & 80.38 & 96 \\
         %\hline
         \textbf{Synthetic} &  & \\
         $\hookrightarrow$ Suno v3.5 & 100.00 & 100 \\
         $\hookrightarrow$ \textit{Suno v3}\textsuperscript{$\dagger$} & 100.00 & 96 \\
         $\hookrightarrow$ \textit{Suno v2}\textsuperscript{$\dagger$} & 99.90 & 78 \\
         $\hookrightarrow$ Udio 130 & 100.00 & 100 \\
         $\hookrightarrow$ \textit{Udio 32}\textsuperscript{$\dagger$} & 39.83 & 96 \\
         \hline
    \end{tabular}
    \caption{Test detection scores (\%) on closed-source models. We include the breakdown for Suno and Udio. $\dagger$ indicate versions unseen during training.}
    \label{tab:closedsource}
    \vspace{-0.5em}
\end{table}

\begin{figure}[h!]
  \centering
  \includegraphics[width=\linewidth]{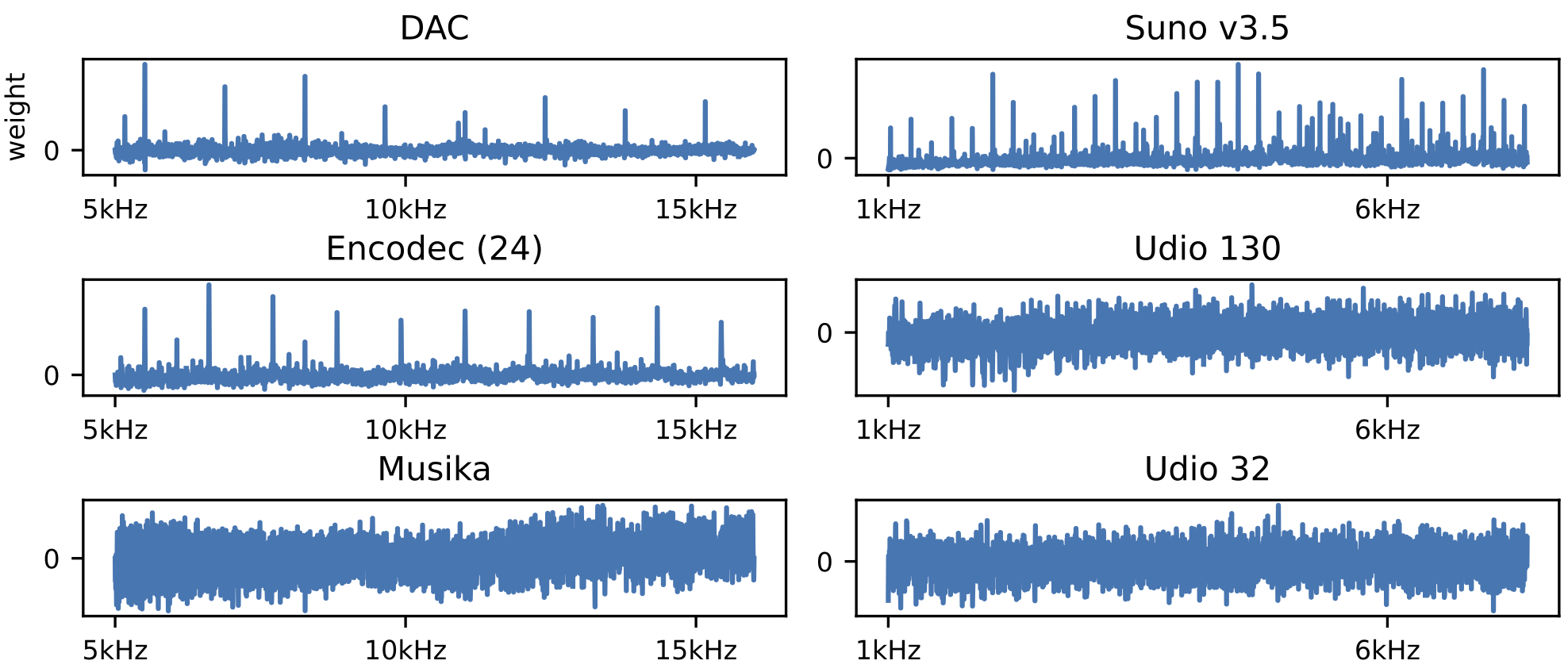}
  \caption{Learned logistic regression coefficients.}
  \label{fig:weights}
  \vspace{-1em}
\end{figure}

We may also train our model to detect each generator individually and display the learned weights as a way to highlight the found peaks (see Figure \ref{fig:weights}). We can see clear patterns and various peak placements from DAC, Encodec, and Suno. This is more fuzzy for other models that are yet still well detected with our method. We leave these more challenging interpretations for future work.

\section{Conclusion}

We propose a theoretical analysis for the emergence of peak artifacts in AI-music, formalize their recursive shape, and predict their architecture independence. Our experiments confirmed this latter observation, and we have used our new-found knowledge to craft a simple detector with performances on par with previous million-parameter models. While this highlights what cues detectors may rely on, other types of artifacts might still exist: \eg this may explain how \textit{SpecTTTra} generalizes to the unseen \textit{Udio 32}.
We also have not touched on the topic of audio manipulation robustness. We can already anticipate that those impacting frequency positions will affect our model (\eg resampling, pitch shift). We leave this as future work.

%To nuance what we said in the introduction, in this section we reveal one kind of artifact that AI-music generator produce. Of course, there is a big difference between showing that one kind of artifact exists with given properties, and telling that all possible artifacts have those. Without further investigation, it is also hard to say whether any AI detector would rely on those artifacts or not or a combination of those with others. Some arguments seem to point in this direction: the artifacts are easy to detect and maybe be so with a linear regression, which — by definition — is a one-layer neural network. It is also known that neural network optimization solutions are attracted towards simple cues, which our artifact seem to be. Therefore, we can only claim that it seems likely and reasonable that deep model would pick up on those, but further interpretability counterfactual analyses would be needed to confirm this for each model: \eg counterfactual analysis where these artifacts are removed, which is not straight-forward to do and requires future work.

\clearpage

% \section{Ethics Statement}

% Very ethical.

% For BibTeX users:
\bibliography{main}

\end{document}